\documentclass{aa}
\usepackage{graphicx}
\usepackage{aalongtable}
\usepackage{txfonts}
\usepackage[]{natbib}
\usepackage{multirow}
\def\lunits{$\rm erg\,s^{-1}$~}
\def\funits{$\rm erg\,cm^{-2}\,s^{-1}$~}
\def\cunits{$\rm cm^{-2}~$}

\def\chandra{{\it Chandra~}}

\begin{document}
 \title{Searching for mid-IR obscured AGN in the CDFN}


  \titlerunning{Mid-IR obscured AGN}
    \authorrunning{I. Georgantopoulos et al.}

   \author{I. Georgantopoulos\inst{1},
           A. Georgakakis \inst{2},
           M. Rowan-Robinson \inst{2},
           E. Rovilos \inst{1} \inst{3}
           }

   \offprints{I. Georgantopoulos, \email{ig@astro.noa.gr}}

   \institute{Institute of Astronomy \& Astrophysics,
              National Observatory of Athens, 
 	      Palaia Penteli, 15236, Athens, Greece \\
              \and
         Astrophysics Group, Blackett Laboratory, Imperial College,  
       Prince Consort Road, SW7 2BZ, U.K. \\
              \and 
      Present address: Max Planck Institut f\"{u}r Extraterrestrische Physik, Giessenbachstrasse, 
   D-85748  Garching, Germany 
             }

   \date{Received ; accepted }

\abstract{  The  efficiency  of  mid-infrared  selection  methods  for
finding obscured AGN  is investigated using data in  the \chandra Deep
Field North.  It is shown  that samples of AGN candidates compiled on
the   basis   of   mid-infrared   colours  only   suffer   substantial
contamination from normal galaxies.  X-ray stacking analysis reveals a
soft  mean   X-ray  spectrum   for  these  sources,   consistent  with
$\Gamma\approx2.1$. This  suggests that star-forming  galaxies and not
obscured AGN  dominate the stacked signal.  In  contrast AGN selection
methods  that  combine mid-infrared  with  optical  criteria are  more
successful  in  finding heavily  obscured  AGN  candidates.  A  method
similar to the one  proposed by Fiore et al. (2008)  is adopted to select
extremely  red objects  ($R-[3.6]>3.7$\,mag)  with high  $24\mu m$  to
optical  flux ratio  ($f_{\rm  24\mu m}/f_{R}>1000$).   About 80\%  of
these sources are not detected at X-ray wavelengths.  Stacking the 
X-ray  photons at the  positions of  these sources  shows a  flat mean
X-ray  spectrum ($\Gamma\approx  0.8$), which  suggests  Compton-thick
sources, low-luminosity and moderately obscured 
 ($\rm N_H\sim 8\times10^{22}$ \cunits) AGN,  or a combination
of  the  two.  The  mid-infrared  colours  and  luminosities of  these
sources are consistent with ULIRGs at $z\approx2$, while {\it HST}/ACS 
images, available  for the optically  brighter of these  sources, show
disturbed optical morphologies in many of them.  The evidence above suggests that this
population includes systems in the process of formation.  
\keywords  {X-rays:  general; X-rays:  diffuse  emission;
X-rays: galaxies; Infrared: galaxies} 
}
   \maketitle
%

\section{Introduction} 

A major challenge  in the quest to constrain  the accretion history of
the Universe is to compile  unbiased samples of active galactic nuclei
(AGN).  Obscuration in  the vicinity  of the  supermassive  black hole
(SBH) and dilution  of the AGN emission from  stellar light associated
with the host galaxy (e.g.  Moran et al. 2002, 
 Severgnini et al.  2003), introduce severe
selection effects  against certain types of  AGN.  X-ray observations,
especially at energies above about  2\,keV, have been shown to be 
efficient  in minimising  these biases.   Hard X-rays,  2-10\,keV, can
penetrate high  column densities  of gas and  dust clouds  and suffer
minimal  contamination by emission  from the  host galaxy,  e.g. X-ray
binaries and hot gas (Hornschemeier et al. 2003).

Recent observations  at X-rays  have highlighted the  effectiveness of
this  wavelength regime  in locating  accreting SBHs.   Firstly, X-ray
luminous AGN  have been revealed  at the centres of  apparently normal
galaxies,  with optical  spectra dominated  by stellar  emission (e.g.
Comastri et  al.  2002; Georgantopoulos et  al.  2003; Georgantopoulos
\& Georgakakis 2005).  Secondly,  the deepest X-ray survey to date, the
2\,Ms Chandra Deep  Field North (CDF-N), finds an  AGN density of over
$\sim  5\,000 \,  \rm deg^{-2}$  (Alexander  et al.   2003; Bauer  et
al. 2004),  an order of magnitude higher  than optically selected  
broad-line QSOs,
which  reach   number  densities   of  only  400\,$\rm   deg^{-2}$  at
$R=24$\,mag \citep{Wolf2003}.

The X-ray surveys in the 2-10\,keV  band , although efficient in compiling
nearly unbiased  AGN samples are not  perfect either.  In  the case of
very  high  obscuring   gas  column  densities,  $>10^{24}$\,\cunits,
Compton  scattering  by   bound  electrons  becomes  important.   This
mechanism  blocks photons with  energies below  almost 10\,keV  at the
rest frame.  If  AGN lie  behind such high  columns densities,  one can
only  hope to  detect  them  indirectly in  the  2-10\,keV band.   For
example in the  case of a toroidal geometry  of the obscuring material
X-ray photons are  expected to be reflected off  the inner surface the
torus into  our line of  sight.  This reflected component  however, is
typically about  2\,dex fainter  than the direct  (obscured) power-law
emission. This  causes even the 2-10\,keV surveys  to under-sample the
Compton-thick AGN  population \citep*{Tozzi2006, Georgantopoulos2007}.
A large number  of active SBH are believed to belong  to this class of
heavily  obscured  sources.  Up  to  50\%  of  the Seyfert-2s  at  low
redshift    for     example,    are    Compton-thick    candidates
\citep*{Maiolino1995, Risaliti1999,  Guainazzi2005, Panessa2006}.  
 Moreover, the
peak of the diffuse X-ray  background (XRB) spectrum at 30-40\,keV 
 (Gruber et al. 1999, Frontera et al. 2007, Churazov et al. 2007) can
only be modeled by invoking  a population of Compton-thick AGN, which
are missing from existing 2-10\,keV X-ray surveys (Gilli et al.  2007).   
 The   evidence  above   underlines  the
significance of Compton-thick AGN for understanding both  the XRB and
the accretion on SBHs across the history of the Universe.  The precise
fraction of these heavily obscured sources is still uncertain however,
by up to a factor  of few (e.g.   Gilli et al.  2007).

The  bias  of the  2-10\,keV  X-ray  imaging  surveys against  heavily
obscured  AGN  has  motivated  alternative methods  for  finding  such
systems.  The mid-infrared wavelength regime (mid-IR; $\rm 3-30\,\mu$m)
in particular, has received much attention recently and is proposed as
a  potentially powerful tool  for identifying  deeply buried  AGN. The
optical and  UV photons from  the AGN heat  the dust of  the obscuring
material  around the  central  engine and  are  re-emitted as  thermal
radiation at  the mid-IR.  Therefore,  surveys with the  {\it Spitzer}
infrared  mission provide  an excellent  opportunity to  recover these
heavily obscured sources.  The  disadvantage of mid-IR surveys is that
the AGN  are outnumbered by  star-forming galaxies and  some selection
method  is necessary  to separate  the two  populations.  It  has been
suggested that  mid-IR colour-colour plots can be  used to effectively
isolate  obscured AGN.   The principle  behind this  approach  is that
luminous AGN have power-law  spectral energy distribution (SED) in the
mid-IR,  while galaxies  have characteristic  black-body  spectra that
peak at  about 1.6\,$\mu$m.  The  mid-IR colours of AGN  are therefore
redder  than  those of  galaxies defining a characteristic wedge.    
Different  combinations of  mid-IR
colours  have  been  proposed  to select  AGN.   \citet{Lacy2004}  and
\citet{Hatziminaoglou2005} define their  selection window based on the
mid-IR colours of luminous  high-redshift QSOs identified in the Sloan
Digital  Sky Survey.  
 These techniques were tested against control samples and turned to 
 be particularly efficient when combined with an additional criterion. 
 Lacy et al. (2004) impose a flux larger than 1\,mJy at 8.0 $\mu m $.  
Stern et al. (2005) impose a magnitude limit of R=19 finding that 
their mid-IR colours selection technique provides a 
 90\% efficiency in detecting AGN. 

In  a
complementary approach  \citet{AlonsoHerrero2006} and
\citet{Donley2007} select  sources with power-law SEDs  in the mid-IR.
The X-ray  identification rate of these  sources is about  50\% in the
1\,Ms  Chandra Deep  Field South  \citep{Giacconi2002},  increasing to
$\approx  85\%$ in  the deeper  2\,Ms  CDF-N. Parallel  to the  mid-IR
studies above,  selection methods using  a combination of  IR, optical
and/or  radio   criteria  have  been  developed,   also  claiming  the
identification of  type-2 AGN  (e.g.  Martinez-Sansigre et  al.  2005;
Daddi et al.  2007; Fiore et al.  2008). 

The selection methods above  have been highly influential by providing
samples  of deeply  buried AGN  candidates, some  of which  may  be so
obscured that remain  undetected to the limits of  the deepest current
X-ray surveys.   Even for these  sources however, getting a  handle on
their X-ray  properties is essential, firstly to  control any residual
contamination by  non-AGN and secondly to  estimate their contribution 
to the XRB.  Stacking analysis is a powerful tool to study the mean X-ray
properties  of   sources  below  the  detection   threshold  of  X-ray
observations  (e.g.    Daddi  et  al.   2007;  Fiore   et  al.   2008;
Georgakakis et  al. 2008).  Here we  use X-ray stacking  to assess the
efficiency of  mid-IR based selection methods  in identifying obscured
AGN.  The data are from  the northern field of the Great Observatories
Origins Deep Survey (GOODS; Dickinson et al. 2003).  We adopt $\rm H_O
=  75  \,  km  \,  s^{-1}  \,  Mpc^{-1}$,  $\Omega_{\rm  M}  =  0.3$,\
$\Omega_\Lambda = 0.7$.

\section{The Data}
 
The CDF-N is centred at $\alpha = 12^h 36^m
49^s.4$,  $\delta =  -62^\circ 12^{\prime}  58^{\prime\prime}$ (J2000)
and has been surveyed extensively  over a range of wavelengths by both
ground-based facilities and space  missions. The multiwaveband data in
this  field include  {\it Chandra}  X-ray observations,  {\it Spitzer}
mid-IR photometry, {\it HST}/ACS high resolution optical imaging, deep
optical  photometry and  spectroscopy  from the  largest ground  based
telescopes.

The 2Ms  {\it Chandra} survey of  the CDF-N consists  of 20 individual
ACIS-I (Advanced CCD  Imaging Spectrometer) pointings observed between
1999 and 2002.  The combined observations cover a total area of $447.8
\,  \rm  arcmin^2$ and  provide  the  deepest  X-ray sample  currently
available.    Here   we   use    the   X-ray   source   catalogue   of
\citet{Alexander2003}, which  consists of  503 sources detected  in at
least one of  the seven X-ray spectral bands  defined by these authors
in the  range $0.3-10$\,keV.  The flux  limits in the  $0.5-2$ and the
$2-10$\,keV   bands   are    $2.5\times   10^{-17}$   and   $1.4\times
10^{-16}$\,\funits, respectively.  The Galactic column density towards
the CDF-N is $1.6\times 10^{20}$\,\cunits \citep{Dickey1990}.

The most X-ray sensitive part of the CDF-N has been
observed in the mid-IR by the {\it Spitzer} mission \citep{Werner2004}
as part of  the GOODS.  These observations cover an  area of about $10
\times 16.5\rm  \, arcmin^2$  in the CDF-N  using both the  IRAC (3.6,
4.5,  5.8 and  8.0$\, \rm  \mu m$)  and the  MIPS ($24\,  \rm  \mu m$)
instruments onboard Spitzer.   Here we use the 2nd  data release (DR2)
of  the IRAC  super-deep images  (version 0.30)  and the  interim data
release  (DR1+) of the  MIPS 24$\,  \rm \mu  m$ mosaic  (version 0.36)
provided by the  GOODS team (Dickinson et al.   2003; Dickinson et al.
in  preparation).  Sources  are  detected in  these  images using  the
SExtractor  \citep{Bertin1996} software.  
The completeness limits i.e. the flux where the counts peak before
dropping off because of incompleteness are 0.7, 0.7, 2.8, 3.5 and 
 56 $\rm \mu Jy$ in the 3.6, 4.5, 5.8, 8.0 and 24 $\rm \mu m$ bands 
 respectively. Full  details on  the source
extraction     and     flux     derivation    are     presented     in
\citet{Georgakakis2007}.  Much of the analysis presented here uses the
$\rm 24 \mu m$ selected sample,  which consists of 1619 sources 
 down to a ($3\sigma$) limiting flux of about 15\,$\rm \mu Jy$.

Multi-waveband optical imaging ({\it  BVRIz'}) in the CDF-N region has
been obtained  using the SUBARU 8.2-m  telescope.
 U-band observations have been obtained at the 4-m Kitt Peak Telescope
 while $HK^\prime$ observations have been obtained at the 2.2-m University 
 of Hawaii telescope. All the observations are described in Capak et al. (2004).
 Deep {\it HST} ACS observations in the B,V,I,Z bands are also available 
 (Giavalisco et al. 2004). However, we choose to use the 
 catalogue of Capak et al. (2004) as this is clearly more suitable 
 for the derivation of photometric resdhifts. 
Here,  we  use the  $R$-band  selected  sample,  which contains  47451
sources  down  to  a  limiting  magnitude  of  $R_{\rm  AB}=26.6$\,mag
(5$\sigma$).   These observations  cover about  $\rm 0.2\,  deg^2$ and
extend beyond the GOODS field of view ($\approx \rm 0.05\,deg^2$). The
$\rm  24 \mu  m$ catalogue  is  first cross-correlated  with the  IRAC
$3.6\mu m$ catalogue  using a matching radius of  2\,arcsec.  For $\rm
24 \mu  m$ sources with  $3.6\mu m$ counterparts  (95 per cent  of the
$\rm 24 \mu m$ population), the more accurate $3.6\mu m$ positions are
used  to  search  for  optical  identifications  within  a  radius  of
2\,arcsec.  There are 1409  $\rm 24 \mu m$/optical associations. Given
the search radius of 2\,arcsec  and the distribution of the positional
offsets  between the  optical  and  the mid-IR  sources,  we expect  a
spurious  identification rate  of  about  4.5\% at  the  limit of  the
optical catalogue, $R\approx26.5$\,mag.  A total of 336 out of the 1409
sources are excluded from the  analysis, because they lie either close
to a bright  source, or at the  edge of the surveyed area  or they are
not detected in at least one of the four IRAC bands.

Optical spectroscopy in the CDF-N  and in the GOODS-North is available
from either observations that specifically target the X-ray population
in these fields  (e.g.  Barger et al.  2003, 2005;  Cowie et al. 2004)
or the  Keck Treasury Redshift  Survey (TKRS).  This program  uses the
DEIMOS spectrograph (Faber  et al.  2003) at the  Keck-II telescope to
observe optically  selected galaxies to $R_{AB}  \approx 24.3$\,mag in
the GOODS-North.   The publicly available catalogue  consists of about
1400 secure redshifts, of which about 500 are associated with $\rm 24 \mu m$
sources.

\section{The sample selection}

The  sample of  mid-IR wedge AGN candidates  is compiled  from the
GOODS-North   $\rm   24  \mu   m$   catalogue.    Two  different   AGN
identification methods  are employed.  The  first one is based  on the
IRAC  band colour  criteria of  \citet{Stern2005}.  The  second method
combines optical  and mid-IR wavebands by adopting criteria similar to 
those proposed by Fiore et al. (2008).

\subsection{The mid-IR wedge AGN candidates} 

The mid-IR wedge AGN candidates lie  in the area of the colour-colour diagram
  proposed by Stern  et al.   (2005)\footnote{$([5.8]  - [8.0])  >  0.6$, $([3.6]  -
  [4.5]) > 0.2  ([5.8] - [8.0] )  + 0.18$, and $([3.6] -  [4.5]) > 2.5
  ([5.8] - [8.0]) -3.5$. [3.6], [4.5], [5.8], and [8.0] are the (Vega)
  magnitudes in the respective  IRAC wavebands}. This selection method
is graphically  shown in  Fig.  \ref{fig_stern}, where in the upper left panel 
we plot the X-ray detected sample (173 sources) while in the upper right panel
 we plot all the sources without X-ray detection (896).   
  Only $\rm 24  \mu m$ 
sources with optical  counterparts   and  detections  in  all  4
IRAC  bands  are employed. The sample of mid-IR wedge selected 
 candidate AGN consists of 177 sources, 46 of them with X-ray detection 
 and 131 not individually detected in the X-rays. 

The \citet{Stern2005} criteria are preferred here over alternative AGN
identification methods which use mid-IR colours only, because they are
likely to suffer less contamination from normal galaxies, while at the
same time remain sensitive to obscured AGN (e.g. Barmby et al. 2006).  

For comparison,  we also explore  the X-ray properties of  the sources
outside the  Stern colour wedge.  This  sample has 896 $\rm  24 \mu m$
sources  with optical  identifications and  detections in  all  4 IRAC
bands.

\subsection{The red/optically-faint mid-IR AGN candidates} 

Fiore et al. (2008) proposed  a method which combines optical, near-IR
and  mid-IR data  to identify  AGN.   These authors  used red  optical
colours ($R-K> 4.5$)  in combination with high mid-IR  ($\rm 24\mu m$)
to optical ($R$-band) flux ratio, $f_{24\mu m}/f_{R}>1000$, claiming a
large number of Compton-thick AGN in the CDF-South.

We  use   similar  selection  criteria   to  compile  the   sample  of
red/optically-faint  mid-IR  AGN candidates  in  the CDF-North.   This
consists of 103 $\rm 24 \mu  m$ sources with $\rm 3.6\mu m$ detection,
$f_{24\mu m} / f_{R} > 1000$  and $R_{AB} - m_{3.6} > 3.7$\,mag, where
$m_{3.6}$  is  the magnitude  at  $\rm 3.6\mu  m$  in  the AB  system.
 The Fiore
et  al. (2008)  $R-K>4.5$\,mag colour  cut  has been  replaced by  the
nearly equivalent  $R_{AB}-m_{3.6}>3.7$\,mag.  This cutoff  is adopted
to  select  sources redder  than  the  elliptical  galaxy template  of
Coleman, Wu \& Weedman (1980) at $z\approx 1$.
Optical counterparts are available for 31 of these sources.
Therefore most of our points refer to $R_{AB} - m_{3.6}$
 and $f_{24\mu m} / f_{R}$ lower limits.

\begin{figure*}
\includegraphics[width=8.cm]{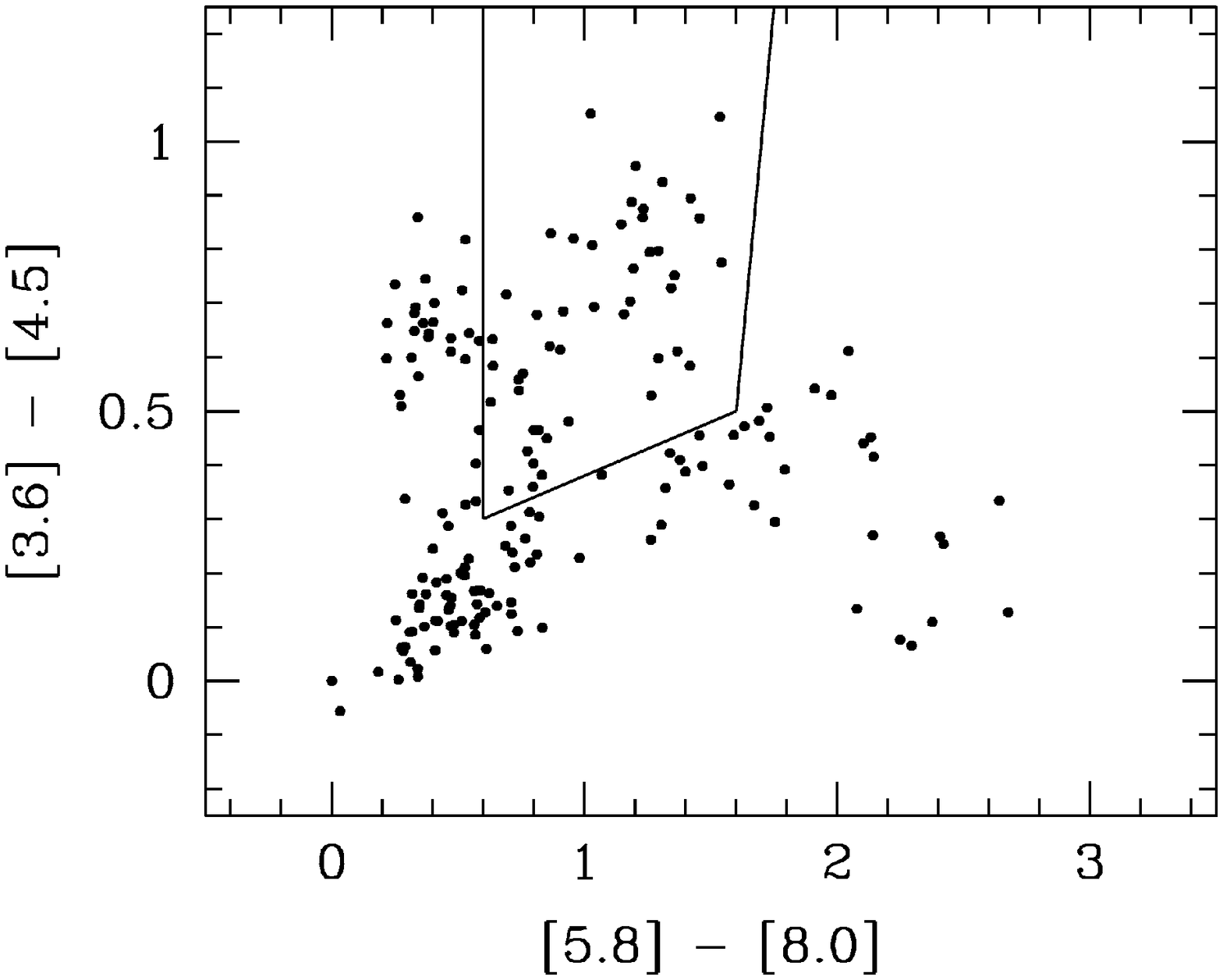}
\includegraphics[width=8.cm]{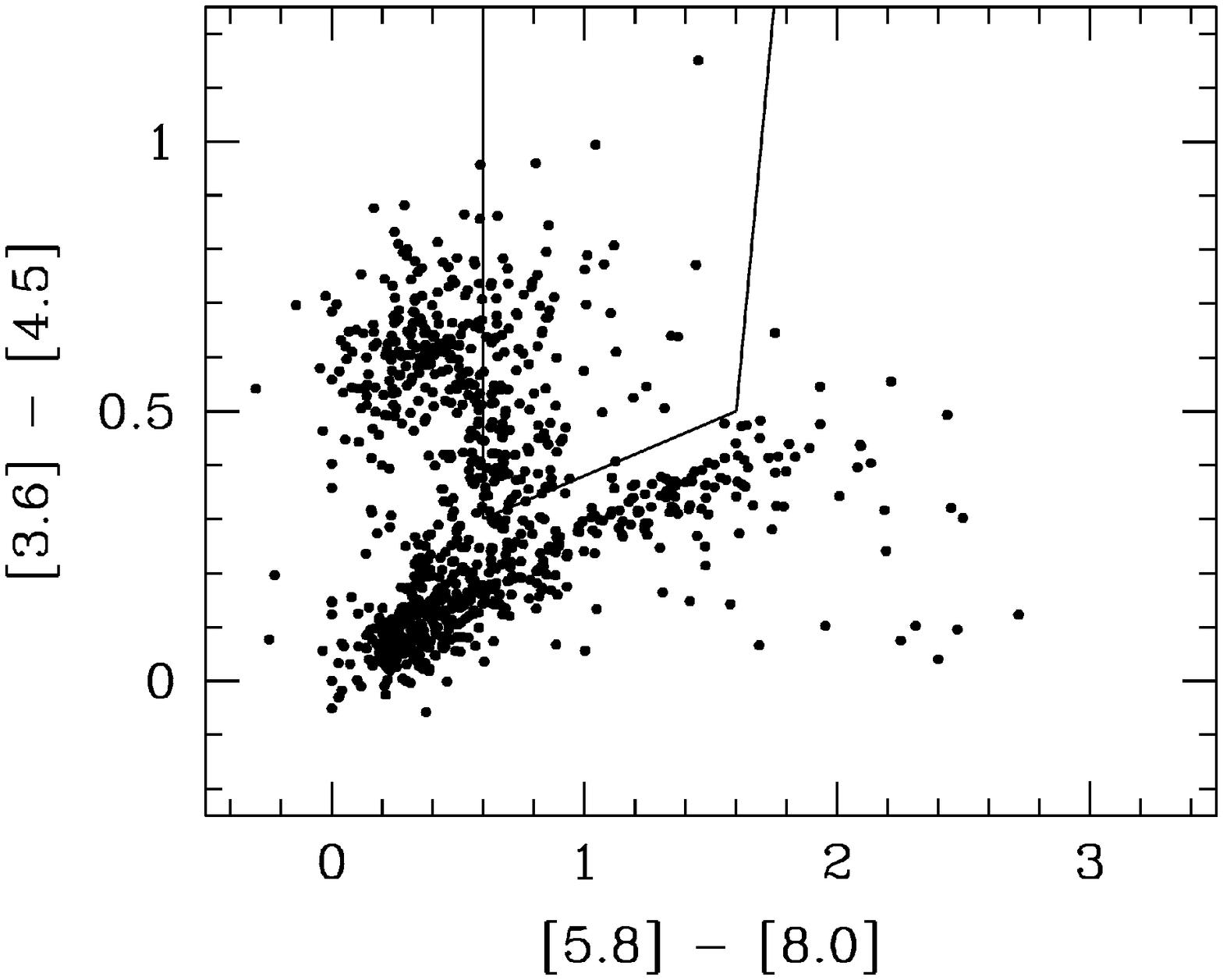}\hfill \\
\includegraphics[width=8.cm]{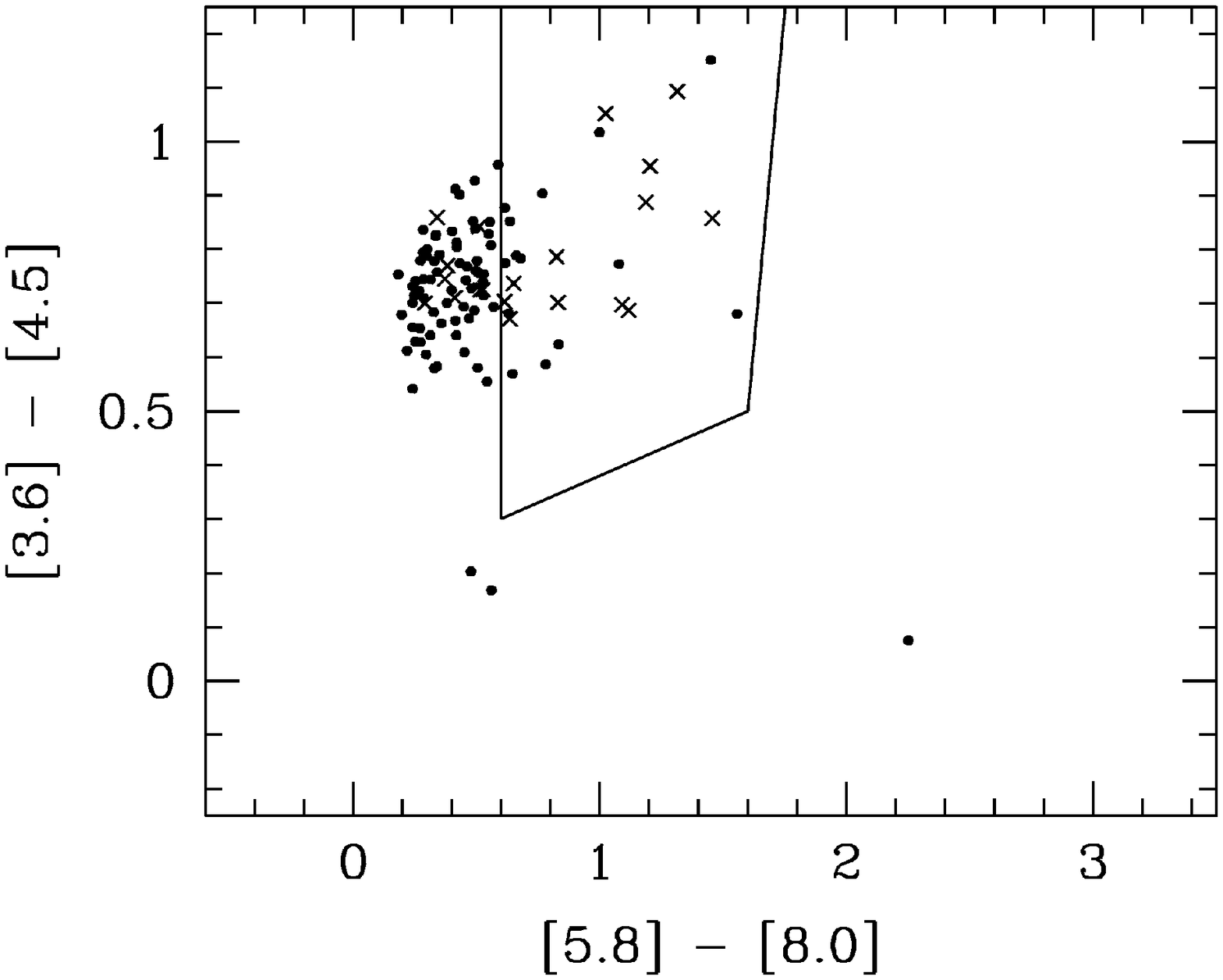}
\includegraphics[width=8.cm]{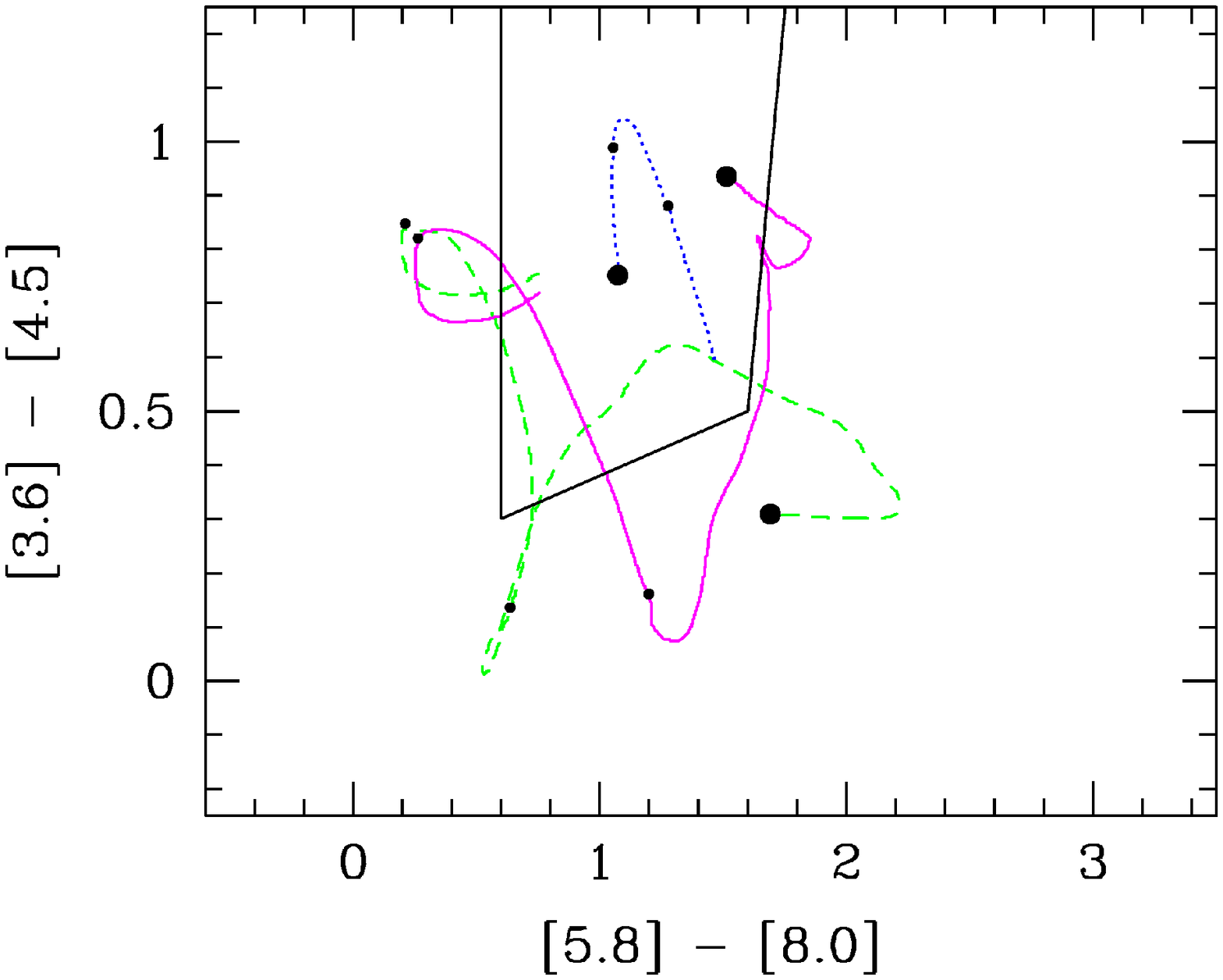}\hfill \\
\caption{Mid-IR colour-colour plot.  For all clarity we use different
panels. In all of them the  wedge defines the AGN region following the
selection criteria  of \citet{Stern2005}. Top left: the $\rm
24\,  \mu  m$  sources  with  X-ray  counterparts; Top right: the  $\rm 24\, \mu
m$ population with detections in all 4 IRAC bands, not detected 
 in X-ray wavelengths; bottom  left:  the
red/optically-faint mid-IR AGN candidates (the crosses and the filled symbols 
 represent the X-ray and non X-ray detections respectively); 
bottom right: colour tracks
for different template  SEDs as function of redshift  up to $z=3$. The
dotted (blue) line represents a  QSO template (Elvis et al. 1994), the
continuous (magenta)  Mrk\,273 (ULIRG, Seyfert 2),  the dashed (green)
Arp\,220  (ULIRG, starburst). For  each template  the position  of the
redshift $z=0$ is marked with a large dot. Small dots correspond 
 to z=1 steps. All templates are adapted from 
Donley et al. (2007)}\label{fig_stern} 
\end{figure*}

The position  of the red/optically-faint mid-IR AGN  candidates on the
Stern et al. (2005) colour diagram is shown in Fig. \ref{fig_stern}.
The majority of  these sources lie outside and on the  left of the AGN
wedge on this  diagram.

\section{The Spectral Energy Distribution fitting method}

For $\rm 24 \mu m$  sources with optical counterparts (1409), we model
the observed UV/optical to  mid-IR Spectral Energy Distribution (SED).
This is  both to determine  photometric redshifts for  sources without
optical  spectroscopy and  to get  an  estimate of  the total  infrared
luminosity.

The  adopted template  fitting  method is  that  of Rowan-Robinson  et
al. (2005).  This is a two  step process.  First, the $U$-band to $\rm
4.5\,\mu m$  photometric data are fit  using a library  of 8 templates
described by Babbedge  et al. (2004), six galaxies  (E, Sab, Sbc, Scd,
Sdm and sb) and two AGN.  One AGN template is based on the SDSS median 
 composite quasar spectrum (Vanden Berk et al. 2001); this is extended 
 to wavelengths longer than 8555A using the IR part of the average QSO
 spectrum defined in Rowan-Robinson et al. (2004). 
 In addition to the composite SDSS template, a simpler template  has 
 been used, based on the mean optical quasar spectrum of Rowan-Robinson 
 (1995) spanning the range 400A to 25 $\rm \mu m$.     
Photometric redshifts are determined at this
stage. The accuracy  is $\delta z/(1+z_{spec})\approx 0.04$ ($1\sigma$
rms), estimated based on the 513 sources with available optical 
spectroscopy.

At  the next  step the  mid-IR SED  of the  sources in  the  sample is
fit. At  longer wavelengths  ($\rm 5.8 -  24 \,  \mu m$) any  dust may
significantly  contribute  or  even  dominate the  observed  emission.
Before fitting models to these wavelengths the stellar contribution is
subtracted  from the  photometric data  by extrapolating  the best-fit
galaxy template  from the previous  step.  The residuals are  then fit
with a mixture of four templates: cirrus (Efstathiou \& Rowan-Robinson
2003),   AGN   dust   tori   (Rowan-Robinson   1995;   Efstathiou   \&
Rowan-Robinson 1995), M\,82 and Arp\,220 starbursts (Efstathiou et al.
2000). The modeling  above provides an estimate of  the total infrared
luminosity, $L_{TOT}$,  in the wavelength range $\rm  3-1000\mu m$. As
discussed  by Rowan-Robinson  et al.   (2005) this  is expected  to be
accurate within a factor of two.

\section{X-ray stacking}

\subsection{Method} 

Stacking techniques have been widely  used in X-ray Astronomy to study
the  mean properties of  source populations  selected to  have certain
well defined  properties and which are  too X-ray faint  to be detected
individually \citep[e.g.][]{Alexander2001,Nandra2002,Georgakakis2003}.

A  fixed radius  aperture is  used  to extract  and to  sum the  X-ray
photons  at  the positions  of  the various  $24  \,  \mu m$  selected
sub-samples. Sources that lie close to or are associated with an X-ray
detection are excluded from the analysis to avoid contamination of the
stacked signal from  the X-ray photons of detected  sources.  We adopt
an  extraction radius  of 3\,arcsec.   This is  found to  maximise the
signal-to-noise  ratio of  the stacked  signal.  A  3\,arcsec aperture
encloses  more than  about 90\%  of  the photons  in the  $1.5-4$\,keV
spectral band at an off-axis angle of 5\,arcmin.  The average off-axis
angle for our sources is 5.5\,arcmin.

The significance  of the  stacked signal depends  on the value  of the
background.   This is  estimated  using the  smoothed background  maps
produced  by the {\sl  WAVDETECT} task  of {\sl  CIAO} by  summing the
X-ray photons in regions around  each source used in the stacking. The
significance of  the stacked signal in  background standard deviations
is  estimated by  $(T-B)/\sqrt{B}$, where  $T$ and  $B$ are  the total
(source + background) and background counts respectively.

An  estimate of  the mean  spectral shape  of the  detected  signal is
obtained  by  performing  the  stacking  in  two  energy  bands,  soft
($0.3-1.5$\,keV) and  hard ($1.5-4.0$\,keV).  Fiore et  al. (2008) has
shown that this choice  of bands minimises the instrumental background
contamination and maximises the  sensitivity of the stacking analysis.
Fluxes are  determined by  multiplying the stacked  count rate  by the
appropriate energy  conversion factor, which  is estimated separately
for each class of sources, based  on the spectral shape of the stacked
signal.


\begin{figure}
\label{fig_z_dist}
\includegraphics[width=9.5cm]{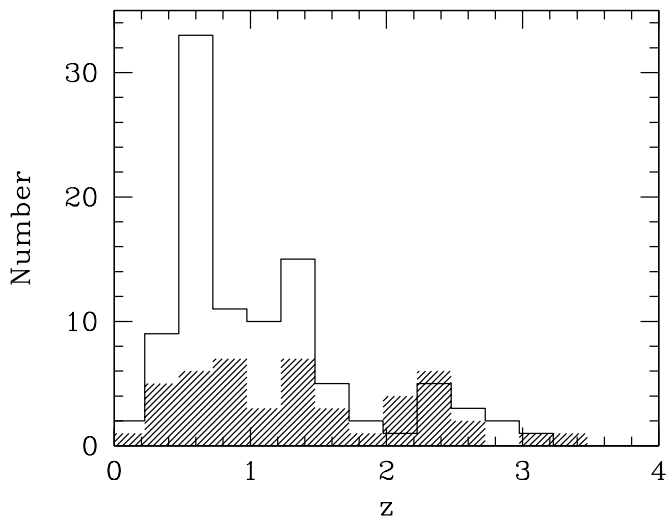}
\caption{The redshift distribution of the 
 mid-IR wedge selected AGN: X-ray detections (shaded) and non X-ray detections 
 (open)  histogram. }
\end{figure}

\begin{figure}
\label{fig_lir_dist}
\includegraphics[width=9.5cm]{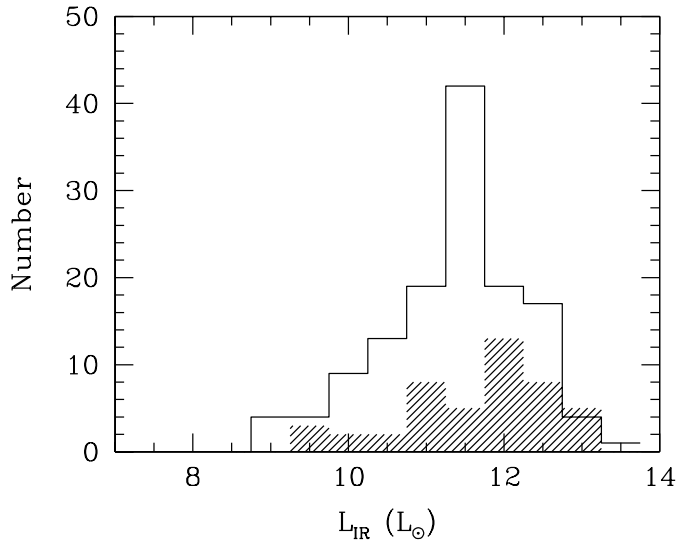}
\caption{The IR luminosity distribution of the  
 mid-IR wedge selected AGN: X-ray detections (shaded) and non X-ray detections 
 (open)  histogram. }
\end{figure}

\begin{table}
\caption{X-ray detected AGN fraction in different mid-IR subsamples}\label{tab_xdet}
\begin{center}
\scriptsize
 \begin{tabular}{l c c c }
\hline
Sample & $N_{TOT}$ &  $N_{X}$ &  $f_X$ \\
       &          &          &  ($\%$)  \\
  (1)  &  (2)     &    (3)   &  (4)   \\
\hline

wedge mid-IR AGN & 177 & 46 & 26 \\
$\rm 24 \mu m $ sources outside & \multirow{2}{*}{896} & \multirow{2}{*}{127} & \multirow{2}{*}{14}\\
wedge \\
red/optically-faint &  \multirow{2}{*}{103} &  \multirow{2}{*}{20} &  \multirow{2}{*}{19} \\
mid-IR AGN\\
\hline
\end{tabular}

\begin{list}{}{}
\item

The columns are:  (1): sample definition. (2): $N_{TOT}$  is the total
number of $\rm 24 \mu m$  sources. (3): $N_{X}$ is the number of X-ray
selected AGN in the sample.  The \citet{Alexander2003} X-ray catalogue
is cross-correlated with  the positions of the $\rm  24 \mu m$ sources
using a matching radius of 3\,arcsec. The X-ray emission of 44 sources
in the  CDF-North are likely dominated  by star-formation (Georgakakis
et al.  2007).  These non-AGN are excluded from the cross-correlation.
(4):  $f_X$ is  the  fraction of  X-ray  detected AGN  in the  sample,
i.e. $f_X=N_{X}/N_{TOT}$. 

\end{list}

\end{center}
\end{table}

\subsection{Mid-IR wedge AGN}
 
Table \ref{tab_xdet} summarises the fraction of X-ray detected AGN in
different  $\rm  24\mu m$  selected  samples.   The  mid-IR wedge  AGN
candidates  have a  relatively high  X-ray identification  rate, about
26\%.  In contrast,  only  about 14\%  of  the $\rm  24\mu m$  sources
outside that wedge  are associated with X-ray AGN.   Fig. 2 shows that
the mid-IR  wedge AGN  candidates lie at  moderately high  redshift on
average,  $z  \approx  1.4$.   The  X-ray  detected  members  of  this
population  have a flatter  redshift distribution  that extends  to $z
\gtrsim 2$.  It appears that  X-ray observations are more efficient in
detecting   the  high   redshift   end  of   the   mid-IR  wedge   AGN
candidates. Fig. 3 plots the total IR luminosity distribution of these
sources.   Most of  the  mid-IR  wedge AGN  candidates  belong to  the
Luminous  Infrared Galaxy class  (LIGs), with  $L_{\rm TOT}  > 10^{11}
L_\odot$.

Mid-IR wedge AGN  candidates that lie close to  or are associated with
an X-ray  detection are excluded  from the stacking analysis  to avoid
contamination of the stacked signal  by the X-ray photons in the wings
of the  Chandra PSF.   The stacking results  for the  remaining mid-IR
wedge  AGN candidates  are  presented in  Table \ref{tab_stack1}.   We
estimate a soft mean X-ray spectrum for these sources, consistent with
$\Gamma \approx 2.1$.  The mean 1.5-4.0 \,keV band X-ray luminosity of
this population  is $6\times  10^{40}$ \lunits. For  comparison, there
are 670 $\rm 24\mu m$ sources  outside the Stern et al. wedge which do
not lie close to an X-ray detection.  Stacking analysis shows that the
mean  hardness ratio  of this  population is  consistent  with $\Gamma
\approx  1.9$  and the  mean  X-ray  luminosity  is $5\times  10^{40}$
\lunits,  i.e. similar  to  that  obtained for  the  mid-IR wedge  AGN
candidates. In passing, we note that the average hardness ratio of the
46 mid-IR  wedge AGN  which are detected  at X-rays  is $-0.09\pm0.01$
corresponding to $\Gamma \sim 1.6$.  

\subsection{Red/optically faint mid-IR AGN candidates} 

The   red/optically-faint  mid-IR   AGN  candidates   have   an  X-ray
identification rate of about 19\% (Table 1). Stacking the
X-ray  photons at  the  positions of  these  83 sources  that are  not
detected  individually at  X-rays yields  a  statistically significant
detection  in  both  the  soft   and  the  hard  energy  bands  (table 3).  
The  mean hardness  ratio corresponds to  a photon
index of  $\Gamma \approx 0.8\pm0.2$.  
However, as the hardness ratios cannot differentiate between 
 an intrinsically flat and an absorbed spectrum, it is 
 equally possible that we are viewing AGN with absorbing column
 densities of $\sim 8\times10^{22}$\, \cunits  for 
an  intrinsic power-law X-ray  spectrum with  $\Gamma=1.9$ and  a mean
redshift of  $z = 2$.   For comparison, the  hardness ratio of  the 20
red/optically-faint  sources with  X-ray  detection is  $0.22\pm0.015$
corresponding  to $\Gamma\sim0.8$.  After  the exclusion  of one  source
which dominates the  hard X-ray signal we obtain  a hardness ratio of
$0.09\pm0.02$  corresponding  to   $\Gamma\sim  1.1$.   Only  a  small
subsample of the red/optically-faint  mid-IR AGN candidates, 31 out of
103,  have optical  counterparts.  The  mean photometric  redshift for
these 31 sources is $z_{photo} \approx 1.6$ and the mean IR luminosity
is  $L_{\rm  TOT}   \approx  10^{12}  L_\odot$,  i.e.   Ultra-Luminous
Infrared   Galaxies  (ULIRGs).   We note that one   of  the   
 optically  identified red/optically-faint  mid-IR AGN  candidates
 is  a sub-mm  source  at a
spectroscopic redshift of $z=2.015$ (Alexander et al. 2005).

\begin{table*}
\begin{center}
\caption{Stacking Results for the wedge selection of Stern et al.}
\label{tab_stack1}
\begin{tabular}{lcccccccccc}
\hline
Type  & No  & $z$ & exposure & Soft counts & Soft Flux & $L_{\rm x}$ & Hard counts  & Hard Flux  & HR & $L_x/L_{IR}$ \\ 
(1)   & (2) & (3) &  (4)     &   (5)       &  (6)      &  (7)       &    (8)       &   (9)      & (10) & (11)  \\ 
\hline
  AGN & 126& 1.44 & 190 & $244\pm37$ (7.1$\sigma$) & $1\times 10^{-17}$ & $6\times 10^{40}$ & $147\pm40$ (3.8$\sigma$) & $7\times10^{-18}$ & $-0.25\pm 0.05$  & $2\times10^{-5}$ \\   
 Galaxies & 670 & 1.05 & 1010 & $1685\pm121$ (15.0$\sigma$)  & $1\times10^{-17}$  & $5\times10^{40}$   & $1136\pm131$ ($8.9\sigma$) & $1\times10^{-17}$ & $-0.19\pm0.02$ & $4\times10^{-5}$  \\  
\hline
\end{tabular}
\begin{list}{}{}
\item 
The columns are: (1) sample definition; (2): Number of sources used in the stacking;  
(3): mean redshift;  (4) effective  exposure time  in Ms;
(5):  net  counts  in   the  0.3-1.5\,keV  band  and  significance; (6):  
flux in the 0.3-1.5\,keV band in
units of \funits; (7): luminosity in the 0.3-1.5\.keV band in units of
\lunits; (8)  net counts  in the 1.5-4\,keV  band and  significance; 
(9)  flux in  the 1.5-4\,keV  band in
units of \funits; (10) Hardness  Ratio; (11): Soft X-ray (1.5-4 keV) to
IR luminosity ratio 
\end{list}

\end{center}
\end{table*}   

\begin{table*}
\begin{center}
\caption{Stacking results for the red/optically faint AGN candidates}
\label{tab_stack2}
\begin{tabular}{lccccccc}
\hline
Type  & No & exposure      & Soft counts   & Soft Flux   & Hard counts  & Hard Flux  & HR   \\ 
(1)   & (2)&    (3)        &  (4)          &  (5)         &   (6)       & (7)        & (8)  \\
\hline
  AGN & 83& 125  & 105$\pm29$ (3.8$\sigma$) & $5.5\times 10^{-18}$ & 164$\pm33$(5.4$\sigma$) & $1.3\times10^{-17}$ & $0.22\pm 0.06$  \\   
 
\hline

\end{tabular}
\begin{list}{}{}
\item The columns  are: (1) sample definition; (2):  Number of sources
used in the  stacking;  (3)) effective exposure time in Ms;
(4):  net  counts  in   the  0.3-1.5\,keV  band  and  significance; 
(5):  flux in the 0.3-1.5\,keV band in
units  of  \funits;  (6)  net   counts  in  the  1.5-4\,keV  band  and
significance; (8) Hardness Ratio.
\end{list}

\end{center}
\end{table*}

\section{Discussion}

\subsection{Mid-IR wedge AGN candidates} 

  The   X-ray   background   population   synthesis   models   of
\citet{Gilli2007} predict a steeply rising number of Compton-thick AGN
just below  the flux limit of  the CDF-N. In  particular, they predict
that  $\sim$25\%   of  the  sources   in  the  $2-10$\,keV   band  are
Compton-thick          AGN           at          fluxes          
$\sim10^{-16}$\,\funits. Extrapolating the AGN number-count distribution
of \citet{Bauer2004} yields about  70 Compton-thick AGN in the common
{\it Spitzer/Chandra} strip at these flux levels.
 At fluxes of $10^{-17}$ \funits, comparable to the 
 flux of the mid-IR wedge AGN, this number rises to 200.  
The application  of the colour  criteria of \citet{Stern2005}  in the
GOODS North sample have identified 131 AGN with no X-ray counterpart.  
The number of mid-IR wedge AGN candidates is of the same 
 order of magnitude as that predicted by the X-ray background 
 synthesis models. 
 We caution that the above is only a very coarse comparison as 
 the XRB synthesis models cannot accurately predict  
 the number of Compton-thick AGN at mid-IR wavelengths.
 This is because some sources can be so obscured that 
 the nuclear flux will emerge only at the longest rest frame wavelengths, 
 outside the IRAC bands, (Silva et al. 2004, Ballantyne \& Papovich 2007). 
 Moreover,  at high redshift ($z>2$)
 the IRAC bands  will sample only the rest frame galaxy emission. 
 

The   steep  mean   X-ray  spectral   properties  of   these  sources,
$\Gamma\approx2.1$,  are  inconsistent with  the  flat spectral  index
expected for Compton-thick AGN  in the {\it Chandra} passband ($\Gamma
\sim 1 $; e.g. Georgantopoulos et  al. 2007).  This makes the case for
a substantial fraction of Compton-thick sources among the mid-IR wedge
AGN  candidates  difficult.  Further  clues  on  the  nature of  these
sources  can be  obtained from  the ratio  between X-ray  and infrared
luminosities. This is estimated  $L_{\rm x}/L_{\rm IR} \approx 2\times
10^{-5}$, approximately 2\,dex lower than unobscured AGN (Elvis et al.
1994),  and comparable to  the ratio  of $L_{\rm  x}/L_{\rm IR}\approx
4\times10^{-5}$, determined  for the mid-IR sources  outside the Stern
et  al.   wedge (see  Table  2).  For  comparison, local  star-forming
galaxies  have $L_{\rm  x}/L_{\rm IR}\approx  8\times10^{-5}$ (adapted
from Ranalli  et al. 2003).  This is  a factor of few  higher than the
value derived for the mid-IR  wedge candidate AGN.  The very low value
of $L_{\rm x}/L_{\rm IR}$ suggests that the mid-IR wedge candidate AGN
sample is largely contaminated  by normal galaxies.  This is demonstrated
in Fig.  1, where the  colour-colour track of the  star-forming galaxy
Arp\,220  overlaps with  the Stern  et  al.  wedge  at specific  broad
redshift intervals.

\subsection{Comparison with mid-IR power-law methods} 

 Mid-IR power-law selection criteria have been proposed to 
 produce AGN samples with a small level of galaxy contamination.
 The mid-IR power-law sources form a subsample of
 the mid-IR wedge sources of Stern et al. (2005).       
 \citet{AlonsoHerrero2006} selected 24\,$\mu$m sources in
the CDF-S which present red power-law SEDs in the {\it Spitzer}
$(3.6-8.0\,\mu$m) bands ($f_\nu \propto \nu^{\alpha}$ with $\alpha<-0.5$). 
 Nearly half of their 92 sources are not detected in
X-rays. \citet{Donley2007} apply similar power-law SED criteria to the Spitzer
sources of CDF-N with the difference that they do not require their sources to
be detected in 24 $\mu$m. They consider 47 AGN within an off-axis
angle of $<$10 arcmin. \citet{Donley2007} find that of these, 30 are detected in
the \citet{Alexander2003} catalogue, 10 more are detected at very low
significance ($>2.5\sigma$) at X-ray wavelengths, and 7 remain undetected. 
 In Fig. 4 we show the position of the 
 power-law selected sources on the mid-IR colour-colour diagram.
 The vast majority fall within the Stern wedge.  
We note that the 17 sources (10 low significance and the seven undetected) 
 of Donley et al. generally avoid 
 the central part of the CDF-N. This suggests that these have not been detected 
 in the \citet{Alexander2003} catalogue simply because the exposure time 
 is lower in these regions due to vignetting.

 In order to compare their work with our results, we estimate the co-added
signal of the 17 sources. 
 The results are presented in Table 4.  We detect a
statistically significant signal in both the soft (0.3-1.5 keV) and the hard 
 (1.5-4 keV) band. The average flux in the 0.5-2 keV  band 
 is $3.4\times 10^{-17}$ \funits, 
 very close to the flux limit of the current CDF-N observation. The
 hardness ratio is $-0.07\pm 0.08$ translating to a photon index of
  $\Gamma\approx 1.55\pm0.15$, or alternatively a column density 
 of $\rm N_H\sim 10^{22}$ \cunits at z=2 (for $\Gamma=1.9$).
 We note that when we stack only the 7 sources which are not 
 detected at X-rays, we obtain no signal in either the soft 
  or the hard band.      
In any case, the number of missed AGN in X-ray surveys
recovered with mid-IR power-law methods is not high. 
\citet{Donley2007} find 7 mid-IR AGN not
detected in X-rays which form a miniscule fraction of the few hundreds X-ray
detected AGN in the same area. This figure is at least one order of magnitude
lower than the estimates of the X-ray background population synthesis models,
which predict large numbers of Compton-thick AGN at fluxes just fainter than
the CDF-N (Gilli et al. 2007). 

\begin{figure}
\includegraphics[width=8.5cm]{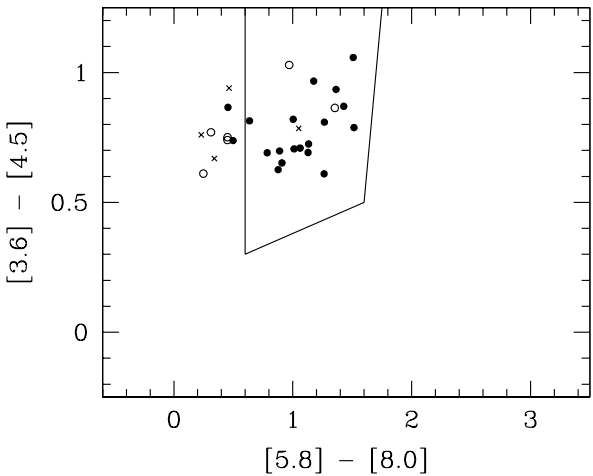} 
\caption{The mid-IR power-law selected 
sources of Donley et al. (2007);
 filled symbols are the X-ray detections, open circles the X-ray non detections and 
 crosses the marginal 2.5 sigma detections (see \S 7.2).}
\end{figure}

\begin{table*}
\begin{center}
\caption{Stacking results for the mid-IR power-law methods}
\label{tab_stack2}
\begin{tabular}{lccccccc}
\hline
Type  & No & exposure      & Soft counts   & Soft Flux   & Hard counts  & Hard Flux  & HR   \\ 
(1)   & (2)&    (3)        &  (4)          &  (5)         &   (6)       & (7)        & (8)  \\
\hline
  AGN & 17& 18  & 78$\pm13$ (7.8$\sigma$) & $ 3.1\times 10^{-17}$ & 68$\pm14$ (5.9$\sigma$) & $3.7\times10^{-17}$ & $-0.07\pm 0.08$  \\   
 
\hline

\end{tabular}
\begin{list}{}{}
\item The columns  are: (1) sample definition; (2):  Number of sources
used in the  stacking;  (3)) effective exposure time in Ms;
(4):  net  counts  in   the  0.3-1.5\,keV  band  and  significance; 
(5):  flux in the 0.3-1.5\,keV band in
units  of  \funits;  (6)  net   counts  in  the  1.5-4\,keV  band  and
significance; (8) Hardness Ratio.
\end{list}

\end{center}
\end{table*}

\subsection{Red/optically faint mid-IR AGN candidates}

Red/optically faint mid-IR AGN candidates are more promising in terms 
of number density for being the Compton-thick population 
predicted by the models of Gilli et al. (2007).  
For the  red/optically-faint mid-IR AGN candidates we  estimate a mean
hardness ratio of $0.22\pm 0.06$,  which is comparable to that derived by
Fiore et al. (2008; $\rm HR\approx0.1$) for similarly selected sources
in   the  CDF-South.    Our hardness ratio corresponds   to  a   photon   index  of
$\Gamma\approx  0.8\pm0.2$,  in the range  of the  spectral  index  of
Compton-thick  AGN which are dominated by the reflected component. 
 The mid-IR--to--X-ray flux ratio  of these sources
is   $\rm    F(24\mu   m)/F(2-8   keV)\sim    6\times   10^{18}   [\mu
Jy]/[erg~cm^{-2}~s^{-1}]$,  similar to that  of the  prototype Compton-
thick  AGN NGC\,1068  at  $z\approx2$  (Daddi et  al.   2007).  It  is
therefore likely that a  large fraction of the red/optically-faint AGN
candidates are associated with Compton-thick sources at high redshift.
Alternatively,  they may  be low-luminosity  AGN obscured  by 
column densities, $N_H\sim 8\times10^{22}$\,\cunits at redshift of z=2.  
 Good quality X-ray spectra could differentiate between the 
above scenaria.  However, the average
flux of these sources is about a factor of four lower than the current
flux  limits of  the 2\,Ms  CDF-N in  both the  0.3-1.5 and  1.5-4 keV
bands.  This means that a total  of 8\,Ms of {\it Chandra}  observations 
are needed just to detect this  population in the soft-band at least in  the 
central part  of the field-of view, where the background will 
remain photon limited (see Alexander et al. 2003).
Alternatively, near-IR spectra could be useful in detecting 
 AGN signatures such as a strong \ion{[OIII]} line. 

The mid-IR  colours of  the red/optically-faint mid-IR  AGN candidates
provide  additional  information on  their  nature.   In  Fig.  1  the
majority of  these sources  lie in the  region of the  parameter space
occupied  by  ULIRGs, such as Arp220 or Mrk273, at  redshifts  
 $z\approx2$,  i.e.   galaxies  in
formation,  possibly through mergers  (Mirabel \& Sanders 1996).
This is consistent with the mean photometric redshift of $\approx 1.6$
and the mean total infrared luminosity of $\approx 10^{12}\L_{\odot}$
estimated for  the 31  red/optically-faint mid-IR AGN  candidates with
optical  counterparts.  High resolution  {\it HST}/ACS  images also
suggest that these sources  include interacting systems.  There are 20
sources bright enough  in the $i$ or $z$  bands for visual inspection.
Many of them  (13) either show disturbed optical morphology
 or are in close pairs/associations. One of them is
associated  with   a  sub-mm   source  at  $z=2.015$.     
Although  the morphological   subsample   represents  only  about  
 20\%   of   the
red/optically-faint  mid-IR  AGN  candidates,  the evidence  above  is
suggestive  that this  population  includes galaxies  captured in  the
process of  formation via tidal interactions.  In  this respect, these
obscured AGN bear many  similarities to high redshift systems believed
to undergo the concurrent growth of their stellar population and their
central SBH.   Both sub-mm galaxies  at $z\approx2$ (Alexander  et al.
2005)  and mid-IR  excess 24$\mu  m$ selected  sources  at $1.4<z<2.5$
(Daddi et  al.  2007) are suggested  to harbour Compton-thick AGN, in
addition to star-formation activity.

Interestingly at  lower redshift, $z\la  1$, most of the  obscured AGN
are  associated  with  early-type  galaxies  rather than systems  in
formation (e.g.  Rovilos \& Georgantopoulos 2007).  Georgakakis et al.
(2008) used stacking analysis to  search of heavily obscured AGN among
different types of galaxies in the range $0.4<z<0.9$.  They showed the
emergence  of an  X-ray faint  and obscured  AGN  population, possibly
including some Compton-thick sources, among galaxies in the red cloud,
where most of the early-type quiescent systems are found. The evidence
above may suggest a change in the properties of the galaxies that host
obscured, possibly Compton-thick,  AGN. At high redshift, $z\approx2$,
an  increasing body  of evidence  points to  a link  with  galaxies in
formation, while at lower redshifts, $z\la 1$, they are found in early
type hosts.

\section{Conclusions}  
 We investigate the X-ray properties of mid-IR selected AGN in the CDF-N,
 using the {\it Spitzer} $24\mu m$ sample. 
 Our conclusions can be summarised as follows.

 \begin{itemize}
 \item{Stacking analysis of the  mid-IR wedge AGN candidates reveals a
soft mean  X-ray spectrum, consistent with $\Gamma\approx  2.1$, and a
low X-ray--to--IR luminosity ratio, both suggesting that the sample is
contaminated by  normal galaxies. We  conclude that the  mid-IR colour
'wedge' methods  fail to  easily identify AGN,  when applied  to faint
optical magnitudes.}

 \item{Mid-IR 'power-law' selection  techniques are more successful in
finding AGN.  These sources overlap substantially  with X-ray selected
AGN   however,    and   do   not   represent   a    new   X-ray   weak
population. Moreover,  the total number  of mid-IR 'power-law'  AGN in
the  CDF-N  is  well below  the  total  number  of Compton-thick  AGN
predicted by X-ray background population synthesis models.  }

 \item{Sources       with       high      24$\mu       m$--to--optical
     ($f_{24}/f_{opt}>1000$)    flux    ratio    and    red    colours
     ($R_{AB}-m_{3.6}>3.7$)   have   hard   stacked   X-ray   spectrum
     ($\Gamma\approx0.8$),   suggesting  Compton-thick   systems,  Low
     Luminosity AGN with column densities of $8\times10^{22}$ \, \cunits 
     or a combination of 
     the two.  Most  of these sources occupy a  distinct region of the
     colour-colour  mid-IR  diagram  and  lie  outside  of  the  Stern
     wedge. The mid-infrared colours and luminosities of these sources
     are  consistent with  ULIRGs at  $z\approx2$.  For  the optically
     bright  subsample of  this population,  the {\it  HST/ACS} images
     show evidence for interactions, suggesting systems in the process
     of formation.}
 \end{itemize} 

\begin{acknowledgements}
We thank the anonymous referee for numerous comments 
 and suggestions which helped to improve the paper. 
AG has  been   supported  by  funding  from  the  Marie-Curie
Fellowship  grant MEIF-CT-2005-025108. 
 We acknowledge use of {\it Spitzer} data provided by the 
 {\it Spitzer} Science Center. 
 The Chandra data were taken from the Chandra Data Archive 
 at the Chandra X-ray Center.   
\end{acknowledgements}


%
\end{document}